# Surface Topography: Metrology and Properties

**PAPER**

# Contact.engineering—Create, analyze and publish digital surface twins from topography measurements across many scales

Michael C Röttger[1] , Antoine Sanner[1,5] , Luke A Thimons[3] , Till Junge[2] , Abhijeet Gujrati[3], Joseph M Monti[4], Wolfram G Nöhring[1] , Tevis D B Jacobs[3,*] and Lars Pastewka[1,5,*]

[1] Department of Microsystems Engineering, University of Freiburg, 79110 Freiburg, Germany
[2] Laboratory for Multiscale Mechanics Modeling, EPFL, CH-1015 Lausanne, Switzerland
[3] Department of Mechanical Engineering and Materials Science, University of Pittsburgh, Pittsburgh, PA 15261, United States of America
[4] Department of Physics and Astronomy, Johns Hopkins University, Baltimore, MD 21218, United States of America
[5] Cluster of Excellence *liv*MatS, Freiburg Center for Interactive Materials and Bioinspired Technologies, University of Freiburg, 79110 Freiburg, Germany
* Authors to whom any correspondence should be addressed.

E-mail: tjacobs@pitt.edu and lars.pastewka@imtek.uni-freiburg.de

## Abstract
The optimization of surface finish to improve performance, such as adhesion, friction, wear, fatigue life, or interfacial transport, occurs largely through trial and error, despite significant advancements in the relevant science. There are three central challenges that account for this disconnect: (1) the challenge of integration of many different types of measurement for the same surface to capture the multi-scale nature of roughness; (2) the technical complexity of implementing spectral analysis methods, and of applying mechanical or numerical models to describe surface performance; (3) a lack of consistency between researchers and industries in how surfaces are measured, quantified, and communicated. Here we present a freely-available internet-based application (available at https://contact.engineering) which attempts to overcome all three challenges. First, the application enables the user to upload many different topography measurements taken from a single surface, including using different techniques, and then integrates all of them together to create a digital surface twin. Second, the application calculates many of the commonly used topography metrics, such as root-mean-square parameters, power spectral density (PSD), and autocorrelation function (ACF), as well as implementing analytical and numerical calculations, such as boundary element modeling (BEM) for elastic and plastic deformation. Third, the application serves as a repository for users to securely store surfaces, and if they choose, to share these with collaborators or even publish them (with a digital object identifier) for all to access. The primary goal of this application is to enable researchers and manufacturers to quickly and easily apply cutting-edge tools for the characterization and properties-modeling of real-world surfaces. An additional goal is to advance the use of open-science principles in surface engineering by providing a FAIR database where researchers can choose to publish surface measurements for all to use.

## 1. Introduction: The challenges impeding a scientific and practical understanding of roughness-dependent surface properties

Surface topography controls the function of material interfaces [1]. More than 100 years ago, Binder [2] showed that the electrical resistance between two flat surfaces in contact was far higher than would be expected given the material properties and overall geometry of the contact. In the following years, it became clear that, because of surface roughness, the area of true contact between materials is typically far lower than the apparent area of their macroscopic contact [3, 4]. Since then it has been shown that virtually all engineering materials, even the most highly polished ones, have surface roughness over some range of length scales [5]. For emerging manufacturing approaches, such as additive manufacturing (3D printing) the effect of surface finish can be even more dramatic, with severe consequences e.g. for fatigue [6]. However, whether mature or advanced manufacturing techniques are used, this surface roughness significantly alters surface properties—including adhesion, friction, lubrication, elastic and plastic deformation under load, electrical transport, and thermal transport [7–10].

Over the last sixty years, a wide variety of analytical and numerical models have been proposed to describe roughness-dependent properties [3, 11–16], as described in Sect. 5. Especially in the last two decades, significant advances have been made in accounting for the multi-scale nature of roughness. However, there remains no consensus in the scientific community about which models best describe real surfaces under which conditions. Even the most effective models have only limited application in industrial contexts to improve product performance. To date, the modification of surface finish is mostly empirical, and it remains impossible to rationally design the optimal surface topography to precisely tune a given surface property [17–19].

The disconnect between scientific advancement of roughness theories and experimental validation and use of these theories arises because of three central challenges, as described below.

*Central challenge 1: The difficulty of integrating different surface measurements to capture the multi-scale nature of roughness.* Surface roughness is known to exist at many length scales and therefore cannot be fully described from any single measurement, such as a line scan from a stylus profilometer or a topographic map from an atomic force microscope (AFM). Instead, a comprehensive description requires many surface measurements across a wide range of magnifications, including using different instruments and techniques. However, it is challenging to fuse these disparate measurements into one comprehensive description of a surface, because of their different size scales, resolutions, collection/analysis software, and file formats.

*Central challenge 2: The technical complexity of implementing spectral analysis methods, and of applying mechanical or numerical models to describe surface performance:* Spectral analysis methods can be complex, often relying on signal processing tools, while technical implementation choices affect the precise value of the outcome. (See, for example, the wide array of calculations and units that are commonly used for the power spectral density [20].) Furthermore, analytical and numerical models of contact tend to be even more challenging to implement, often requiring complex calculations or even numerical solvers that are not accessible for typical experimentalists in either research or manufacturing contexts. This technical complexity and the ambiguity surrounding specific implementations present a significant barrier to the validation and use of roughness models to real-world surfaces.

*Central challenge 3: A lack of consistency between researchers and industries in how surfaces are measured, quantified, and communicated:* The characterization and specification of surface roughness varies enormously. Manufacturing contexts favor simple scalar parameters that are easy to measure and that are specified clearly using reference standards (e.g., ISO 4287 and ASME B46.1), yet there is not even agreement about which of the dozens of scalar parameters should be measured and specified ($R_a$, $R_q$, $R_z$, $R_{\Delta q}$ etc.). In the context of scientific research, more comprehensive descriptions are preferred such as the power spectral density and the autocorrelation function, but there is a lack of agreement about which is most meaningful. Furthermore, the very act of sharing surface data (with collaborators, or publishing it with a scientific paper) is made complicated by proprietary software applications with varying file formats, and a lack of agreement about how to save and report raw data.

These three central challenges cause a gap in the current understanding of topography, which must be filled in order to predict and control roughness-dependent properties. The remainder of the paper describes a software platform that attempts to close this gap.

## 2. Contact.engineering: overview and guiding principles

In order to address all three of the central challenges discussed in the previous section, we have created a freely-available internet-based application for rough-surface analysis that can be accessed at https://contact.engineering/. The source code behind this application is freely available[6] such that power-users can directly use the computational engine for surface topography analysis in their workflows. The present paper describes the choices made for the web application that is accessible at *contact.engineering*. These choices are implemented as defaults in the computational engine and serve to set a standard for topography analysis. Users of the back-end Python code can deviate from these choices by overriding these defaults. The application and back-end code are designed around functionalities and guiding principles that address each of the central challenges laid out in the previous section (see figure 1).

We note that a variety of software tools for handling topography data already exist. A popular tool is the open-source software *Gwyddion* [21] that is used by many researchers around the world. Commercial offerings with related functionality include *OmniSurf*

---

[6] See https://github.com/ContactEngineering/.

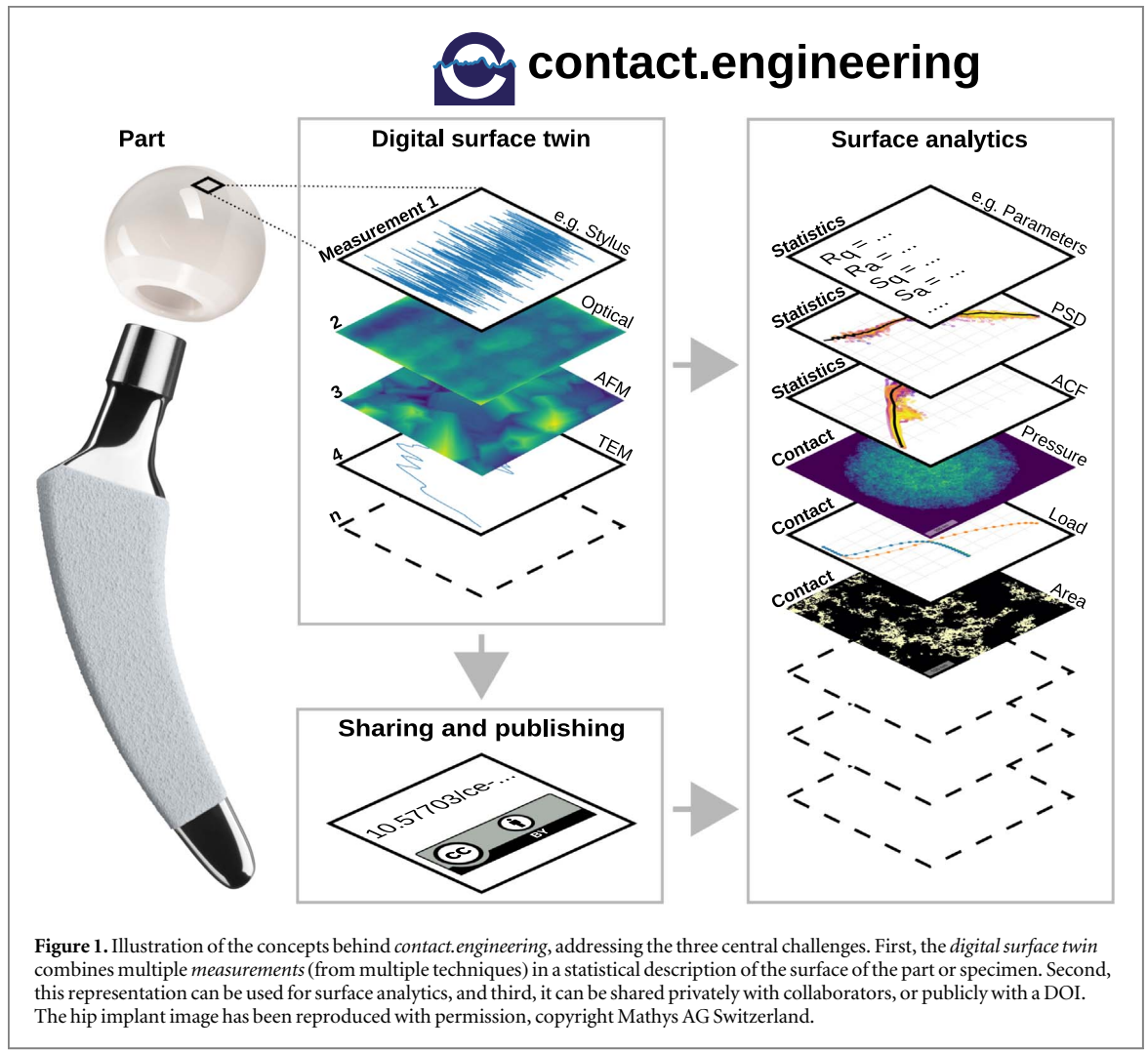

**Figure 1.** Illustration of the concepts behind *contact.engineering*, addressing the three central challenges. First, the *digital surface twin* combines multiple *measurements* (from multiple techniques) in a statistical description of the surface of the part or specimen. Second, this representation can be used for surface analytics, and third, it can be shared privately with collaborators, or publicly with a DOI. The hip implant image has been reproduced with permission, copyright Mathys AG Switzerland.

[22] and *MountainsMap* [23] or software developed by metrology manufacturers and shipped with instruments (e.g. *Talysurf*). While each of these excellent tools serves an important purpose in characterizing topography, there are remaining gaps that these are not designed to address. For instance, while some of the existing tools can do very sophisticated correlative microscopy where the same region of interest is characterized with multiple techniques (e.g. optical and scanning electron microscopy), this is a different functionality than combining many different types of measurements in many locations for the statistical characterization of a material's surface (Challenge 1). Additionally, without the integration of information from multiple different scales and sources, it is difficult or impossible to accurately predict functional properties of surfaces (Challenge 2). Finally, it is common for the source code of commercial software, and thus the implementation of mathematical algorithms, to be hidden from the user. Therefore, these tools cannot be used for fulfilling reproducibility requirements of scientific research [24]. The aim of *contact.engineering* is to address the remaining gaps left by current offerings.

### 2.1. Addressing central challenge 1: The creation of *digital surface twins*

The application allows the user to stitch together many individual measurements of the same surface, such that many limited-size-scale measurements can be combined into a complete statistical representation of the whole surface, called a *digital surface twin*.

Many instruments exist for measuring surface topography. The most popular instruments use either optical methods (white light interferometry, confocal microscopy, and others) or contact-based techniques (atomic force microscopy, stylus profilometry, and others). Any individual measurement of a surface is inherently incomplete, limited in scan size and resolution, and subject to any artifacts of that method. This raw topography data is uploaded to the application as a *measurement*.

Many different *measurements* are collected in a *digital surface twin*, which is an abstract representation of the surface of a physical object. We have implemented analysis tools (described below) that can compute properties of the digital surface twin by combining the individual properties of the *measurements*. While a single topographic measurement can never be representative of the full surface, a collection of many measurements can comprise a comprehensive statistical description, especially if the same surface is measured many times with many different instruments at various locations, scan sizes, and resolutions. The main idea behind *contact.engineering* is to provide a natural way of combining many measurements for the comprehensive description of the surface topography of a specimen, component, or system.

We call this collection of measurements a digital surface twin to reflect the intent, which is to create a virtual model designed to accurately reflect a physical object, i.e. the surface. The virtual version is intended to be a construct of a wide variety of independent measurements, and to enable calculations and simulations to be performed on the construct which could not be accurately performed from the independent measurements in isolation. In the more general context of manufacturing, this digital surface twin would be a part of the full digital twin of a component or system. In the context of scientific research, the digital surface twin is useful to analyze the functional properties of a surface and their dependence on the multi-scale topography.

### 2.2. Addressing central challenge 2: Open-source digital workflows for statistical analysis of surface topography and the prediction of properties

The data uploaded to *contact.engineering* are analyzed through a set of digital analysis workflows. There is a continuously growing number of available analysis workflows, for example for statistical analysis or contact mechanics. These workflows can operate on *measurements*, *digital surface twins*, or both, and produce results that are displayed to the user through the web application. The currently available workflows are described in this paper.

Briefly, the application enables statistical analysis to describe surfaces (described in Sect. 4) for the calculation of mathematically consistent scale-dependent statistical descriptors, such as the power spectral density [20], the height-difference autocorrelation function [25], the variable bandwidth method [26–30], and the newly established scale-dependent roughness parameters (SDRP) analyses [31]. Additionally, the application performs mechanics calculations to compute contact properties (described in Sect. 5), such as the contact area, contact stress, and contact stiffness as a function of load, through the application of numerical models of continuum mechanics. All computational workflows are composed of open-source software.

All of the available analysis workflows are automatically run on each uploaded topography. Some of these have parameters that can be changed by the user, which results in the re-running of the workflow with the new parameters. The database stores *all* workflow results such that requesting a calculation with identical parameters will immediately yield the result already stored in the database. The idea is that results are available to the user at the click of a button without incurring long delays, even for those workflows that require significant computational effort.

### 2.3. Addressing central challenge 3: Enabling (optional) data sharing, fulfilling open-data and FAIR principles

Digital surface twins are securely stored on *contact.engineering*, but enable the user to choose to *share* the surfaces and analyses with colleagues and collaborators, or to *publish* them for the general public, for example, as a companion to a journal article. This publishing of digital surface twins generates a digital object identifier (DOI) [32] that points back to the full collection of measurements that comprises the digital surface twin. Attached to these published digital surface twins are the workflow results described above, such that not only the underlying raw data is available but also a set of derived properties. Once logged in, a user can individualize the parameters of the workflows of these published surfaces and hence directly reuse this data for future scientific work.

Users can log into the system using an ORCID [33] identifier (which anyone can create), or can access a limited functionality without logging in. As with scientific publications, the ORCID identifier of the owner of a digital surface twin is attached to the published versions, and that owner has responsibility for the correctness of the data. Co-authors can be specified upon publication of the data set.

Anyone can search the database of published digital surface twins and download the underlying raw data from the repository. The download contains the original (as-uploaded) dataset as well as a normalized variant that uses the binary, self-describing *NetCDF* format [34] that represents the measured data after potential detrending and filling of missing values (see Sect. 3.2). *NetCDF* is directly readable with common mathematical software tools such as *MATLAB* or *Python/numpy*. This conversion of vendor-specific data formats to *NetCDF* hence enables subsequent reuse of the data in custom analysis scripts. In this way, the website aids the FAIR principles of data handling [35]—making data findable, accessible, interoperable, and reusable. In particular in tribology, solutions for FAIR data are presently lacking [36]. To date no commercial or open-source software solution offers the possibility to share and publish topography data according to these principles; *contact.engineering* provides this needed capability.

## 3. Data handling: The creation, sharing, and publishing of digital surface twins

### 3.1. Data formats
A digital surface twin can be created by uploading a collection of measurements from a single physical specimen (see figure 1). Individual measurements can take various forms, and derive from a wide variety of different measurement approaches. Many topography instruments, including optical profilometry and scanning probe microscopy, measure the height $h$ over a square area. These are referred to as *area scans.* These measurements are therefore described by a function $h(x, y)$, where $x$ and $y$ describe the position in the plane of the surface. The lateral ordinates $x$ and $y$ run from 0 to $L_x$ or $L_y$, respectively, the size of the topographic map. Topographies are exclusively stored on a *regular rectilinear grid* such that the function $h(x, y)$ is discretized into $N_x \times N_y$ equally spaced points, $h_{m,n} = h(m\Delta x, n\Delta y)$, with grid spacing $\Delta x$ and $\Delta y$, $L_x = N_x \Delta x$ and $L_y = N_y \Delta y$ where $m \in [0, N_x - 1]$ and $n \in [0, N_y - 1]$.

Other topography measurements, such as many stylus profilometers, yield the height of the sample along a line profile, and are called *line scans.* Most commonly, line scans can be stored on regular grids such that $h_m = h(m\Delta x)$. However, an alternative way of producing line scans is to take an optical- or electron-microscope image of a cross-section or side-view of a material, and then to use edge-finding or other image analysis to extract its surface contour [37, 38]. This typically yields line scans on a nonuniform grid, $h_m = h(x_m)$, but with irregularly spaced $x_m$.

A variety of data formats exists for area scans, and *contact.engineering* supports reading several native formats (e.g., with file suffixes .di, .spm, .opd, .opdx, .zon, and many others). When reading height data from a native format, we extract additional metadata from these files such as the date of the measurement or the name of the instrument, so that this information does not need to be provided manually by the user. As the application continues to develop, we will further expand the number of supported data formats and the amount of metadata that is automatically extracted. For any unsupported format, or unconventional approach (such as cross-sectioning and contour digitization), topography data can be uploaded using generic formats. As generic data formats for height data, we support plain text, *MATLAB*, *numpy*, *NetCDF* and *HDF5* files. Plain text files are parsed for metadata (header) information and we support standardized metadata as for example provided by the text export function in *Gwyddion* [21]. Line scans on irregular grids can only be uploaded in text format where the text file contains rows of (x, y)-coordinates separated by a space.

A specific feature of irregular grids is that reentrant surfaces can be represented, i.e. those for which there are cliffs or overhangs leading to more than one value of height for a single lateral position. Some methods, in particular cross-sectioning and contour digitization, lead to a discretization of the surface in terms of a set of points $(x_k, h_k)$ connected by straight lines (linear interpolation) that describe the bounding surface. A description of a surface in terms of a discretized curve can therefore be incompatible with a description in terms of a function $h(x)$ where a single position $x$ is assigned a single height value. The web application supports line scans on nonuniform grids but has only limited support for reentrant surfaces at present.

### 3.2. Data preprocessing

Experimental measurements often need to be purified from instrument artifacts before performing analyses. *contact.engineering* implements some basic preprocessing filters to remove such artifacts, so that preprocessing parameters are documented and editable and the original data is stored.

Sample tilt and artificial curvature from the tool can be removed by subtracting the first- or second-order polynomial that minimizes the RMS height of the detrended profile (least-squares fit). Otherwise the mean value of the heights is set to 0.

When using optical profilometers, it is common that individual patches of the surface are left undefined. While contact calculations can still be performed leaving these values undefined, computing the power-spectrum or autocorrelation function requires the interpolation of the missing values. Our method to choose these values is to minimize the RMS slope of the resulting topography, which implies filling each island of missing values (patch) using a harmonic function. The harmonic function is constructed by solving the Laplace equation with the values on the edge of the patch as Dirichlet boundary conditions. This construction leads to exact filling for linear fields and has the property that the mean of the interpolated field is equal to the mean of the height over the edge (by virtue of the mean value property of harmonic functions). For line scans, this is simply a linear interpolation of missing data. For area scans, this leads to well-behaved interpolations without jumps even if large patches are missing.

### 3.3. Data sharing

Once created, the digital surface twin, along with all corresponding analysis (see Sect. 4 and 5), is securely stored on the database. The digital surface twins and their analysis can also be readily shared with colleagues and collaborators within the same institution or across the world. The owner(s) of the topography control access rights (such as view-only, editing, etc.). Additionally, the digital surface twin can be published, such that anyone can access the data for viewing, downloading, verifying, reusing, or re-running *contact.engineering* workflows. Publication of a digital surface twin triggers the creation of a DOI, which links to the twin. All analyses of the twin and individual measurements can be reached from the DOI landing page and retriggered with user-defined parameters. These DOIs can be referenced in publications, facilitating the goal of open data in the scientific community, and can be used for satisfying data availability requirements of funding agencies. This publication function could even be used by a manufacturer to demonstrate a high-performance surface finish of a novel coating or manufacturing process to the scientific community or a customer.

## 4. Statistical analysis: Scalar parameters and correlation functions

The web application displays the uploaded topography and also computes scalar roughness parameters, scale-dependent roughness parameters [31], the power spectral density [20], the height-difference autocorrelation function [25], and performs a variable bandwidth analysis [26–30]. These methods are described briefly below, including the specific algorithms used and recommendations for their use in the context of surface metrology.

### 4.1. Spectral analysis using a Fourier representation

To generate synthetic topographies [20, 39], to measure Hurst exponents [29, 40], or to formulate contact theories [12, 13], it is often convenient to work with a spectral (Fourier) representation $\tilde{h}(q, k)$ of the topography rather than the topography $h(x,y)$ itself. The Fourier representation is required to compute the PSD as described in more detail below. In computing the Fourier transform, we follow the conventions laid out in [20]. The discrete Fourier transform (DFT) of the discrete topography map $h_{m,n}$ is given by

$$\tilde{h}_{o,p} = \Delta x \Delta y \sum_{m,n} h_{m,n} e^{-i(q_o x_m + k_p y_n)}, \quad (1)$$

with pixel size $\Delta x = L_x/N_x$ and $\Delta y = L_y/N_y$. The wavevector $\vec{q}_{o,p} = (q_o, k_p)$, with $q_o = 2\pi o/L_x$ and $k_p = 2\pi p/L_y$. The inverse DFT is then given by

$$h_{m,n} = \frac{1}{L_x L_y} \sum_{o,p} \tilde{h}_{o,p} e^{i(q_o x_m + k_p y_n)} \quad (2)$$

Note that it is important to lay out conventions for the DFT, as different authors use different prefactors in equations (1) and (2). We use a Fast-Fourier-Transform (FFT) algorithm to numerically compute the DFT.

### 4.2. Computing derivatives

Statistical analysis often involves the computation of a local derivative of the heights $h(x, y)$, e.g. $\partial h/\partial x$. For a discrete set of data $h_{x,y}$, there are multiple ways of computing a derivative that lead to different numerical answers. The simplest example is the forward-differences expression for the derivative,

$$\left(\frac{\partial h}{\partial x}\right)_{x_m,y_n} \approx (D_x h)_{m,n} \equiv \frac{h_{m+1,n} - h_{m,n}}{\Delta x} \quad (3)$$

and the central-differences expression for the second derivative

$$\left(\frac{\partial^2 h}{\partial x^2}\right)_{x_m,y_n} \approx (D_{xx} h)_{m,n} \equiv \frac{h_{m+1,n} - 2h_{m,n} + h_{m-1,n}}{\Delta x^2} \quad (4)$$

Higher-order schemes can be systematically derived, for example from finite-difference approximations. Equation (3) is used in the computation of the root-mean-square slope parameter $S_{\Delta q}$ described in Sect. 4.4. More details on such scalar roughness measures are given in the next section.

Within an analysis in terms of the Fourier representation of the topography, the derivative is usually computed by taking the analytical derivative of the Fourier-interpolated function. The result of this construction differs from equation (3) and has the additional problem that it may introduce Gibbs ringing into the solution [41]. However, we can also represent discrete derivatives (of the form given by equations (3) and (4)) using their respective Fourier representations. Inserting equation (2) into (3) yields

$$\begin{aligned}(D_x h)_{m,n} &\equiv \frac{1}{L_x L_y} \sum_{o,p} \frac{e^{iq_o \Delta x} - 1}{\Delta x} \tilde{h}_{o,p} e^{i(q_o x_m + k_p y_n)} \\ &= \frac{1}{L_x L_y} \sum_{o,p} \tilde{D}_x(q_o, k_p) \tilde{h}_{o,p} e^{i(q_o x_m + k_p y_n)}\end{aligned} \quad (5)$$

with

$$\tilde{D}_x(q, k) = \frac{e^{iq\Delta x} - 1}{\Delta x} \quad (6)$$

Here $\tilde{D}_x(q_x, q_y)$ is the Fourier-space representation of the discrete derivative operator $D_x$ given by equation (3). ($D_x$ is an operator while $\tilde{D}_x$ is a complex number.) We can turn any discrete derivative defined in terms of a local 'stencil' into a scalar function such as equation (6). Note that the Fourier derivative (in x-direction) is given by $\tilde{D}_x^{\mathscr{F}}(q, k) = iq$ and that $\tilde{D}_x(q, k) \to D_x^{\mathscr{F}}(q, k)$ for $q\Delta x \ll 1$. The discrete representation of the derivative in not unique but this limiting behavior must be true for any representation. Evaluating derivatives in real-space (via equation (3) or similar) or in Fourier-space (via equation (5)) is numerically *identical*. Within *contact.engineering*, we use a unique representation of the derivative for real-space and Fourier-space analysis. The derivative operators used by *contact.engineering* are summarized in table 1.

### 4.3. Nonperiodic surfaces and windowing

The web application supports both nonperiodic and periodic measurements. Truly periodic surfaces are a rare case that typically only occur in roughness models [20, 39] or computer simulations [42, 43]. Treatment of periodic surfaces is straightforward, as the DFT inherently assumes periodicity. When uploading a topography measurement, the user can declare whether it should be treated as periodic or nonperiodic.

Fourier analysis of nonperiodic surfaces is slightly more complicated. Nonperiodicity introduces ringing artifacts into the Fourier transform that has a signature in the power spectral density $\propto q^2$ and that hides meaningful trends in the data. As is well described in the signal-processing literature, this can be overcome by *windowing*. We define the windowed topography by

$$h_{\text{windowed}}(x, y) = h(x, y) w(x, y) \qquad (7)$$

where $w(x, y)$ is the windowing function. We use a Hann window, but care needs to be taken that the window is spherically symmetric for topography maps and that it is normalized according to $\int w^2(x, y) dx dy = 1$. (See [20] for a detailed discussion.) For periodic surfaces, $w(x, y) = 1$. Some analysis outlined below is carried out on the windowed topography $h_{\text{windowed}}(x, y)$, but for simplicity we will refer to windowed and bare topographies as $h(x, y)$.

**Table 1.** Discrete derivative operators used for surface topography analysis.

| Operator | Real-space representation | Fourier-space representation |
|---|---|---|
| Derivative in *x*-direction | $(D_x h)_{m,n} \equiv \frac{h_{m+1,n} - h_{m,n}}{\Delta x}$ | $\tilde{D}_x = \frac{e^{iq\Delta x} - 1}{\Delta x}$ |
| Derivative in *y*-direction | $(D_y h)_{m,n} \equiv \frac{h_{m,n+1} - h_{m,n}}{\Delta y}$ | $\tilde{D}_y = \frac{e^{ik\Delta y} - 1}{\Delta y}$ |
| Second derivative in *x*-direction | $(D_{xx} h)_{m,n} = \frac{h_{m+1,n} - 2h_{m,n} + h_{m-1,n}}{\Delta x^2}$ | $\tilde{D}_{xx} = \frac{e^{iq\Delta x} - 2 + e^{-iq\Delta x}}{\Delta x^2}$ |
| Second derivative in *y*-direction | $(D_{yy} h)_{m,n} = \frac{h_{m,n+1} - 2h_{m,n} + h_{m,n-1}}{\Delta y^2}$ | $\tilde{D}_{yy} = \frac{e^{ik\Delta y} - 2 + e^{-ik\Delta y}}{\Delta y^2}$ |

**Table 2.** Scalar parameters and their continuous and discrete expressions.

| Name | Symbol | Continuous | Discrete |
|---|---|---|---|
| *Area properties* | | | |
| RMS height | $S_q \approx h_{\text{rms}}$ | $h_{\text{rms}}^2 = \frac{1}{L_x L_y} \int h^2(x,y) \, dx dy$ | $S_q^2 = \frac{1}{N_x N_y} \sum_{m,n} h_{m,n}^2$ |
| RMS gradient | $S_{\Delta q} \approx h'_{\text{rms}}$ | $h'^{2}_{\text{rms}} = \frac{1}{L_x L_y} \int |\nabla h|^2 \, dx dy$ | $S_{\Delta q}^2 = \frac{1}{N_x N_y} \sum_{m,n} [(D_x h)_{m,n}^2 + (D_y h)_{m,n}^2]$ |
| RMS curvature | $S_{\Delta^2 q} \approx h''_{\text{rms}}$ | $h''^{2}_{\text{rms}} = \frac{1}{4 L_x L_y} \int (\nabla^2 h)^2 \, dx dy$ | $S_{\Delta^2 q}^2 = \frac{1}{4 N_x N_y} \sum_{m,n} [(D_{xx} h)_{m,n} + (D_{yy} h)_{m,n}]^2$ |
| *Profile properties* | | | |
| RMS heights | $R_q \approx h_{\text{rms}}$ | $h_{\text{rms}}^2 = \frac{1}{L_x} \int h^2(x) \, dx$ | $R_q^2 = \frac{1}{N_x} \sum_m h_m^2$ |
| RMS slope | $R_{\Delta q} \approx h'_{\text{rms}}$ | $h'^{2}_{\text{rms}} = \frac{1}{L_x} \int \left(\frac{\partial h}{\partial x}\right)^2 dx$ | $R_{\Delta q}^2 = \frac{1}{N_x} \sum_m (D_x h)_m^2$ |
| RMS curvature | $R_{\Delta^2 q} \approx h''_{\text{rms}}$ | $h''^{2}_{\text{rms}} = \frac{1}{L_x} \int \left(\frac{\partial^2 h}{\partial x^2}\right)^2 dx$ | $R_{\Delta^2 q}^2 = \frac{1}{N_x} \sum_m (D_{xx} h)_m^2$ |

### 4.4. Scalar roughness parameters

The web application also displays the distribution of heights, slopes, and curvatures (computed via the respective operators in table 1) in the uploaded measurement. It also computes the root-mean-square values of these quantities, with the appropriate discrete representations for the first and second derivative. Table 2 summarizes these parameters and the symbols we use. One-dimensional (1D, profile) and two-dimensional (2D, area) versions of these numbers are computed and reported. For area scans, we report profile properties in *x*- and *y*-directions. These properties are averages over the respective consecutive line scans in the area scan. Note that for isotropic area scans, the RMS slope the 2D value is larger than the 1D value by a factor of $\sqrt{2}$, i.e., $S_{\Delta q} = \sqrt{2} R_{\Delta q}$. We therefore explicitly call the 2D value $S_{\Delta q}$ *RMS gradient* and not RMS slope to highlight this difference.

Our RMS parameters are similar to *R*- and *S*-parameters of common standards (SEMI MF1811 or ISO 4287) but we would like to point out a couple of differences. First, ISO 16610 recommends the application of two Gaussian filters: one small-wavelength filter to remove instrument noise; and one large-wavelength filter to distinguish 'roughness' from 'waviness'. Here we do not apply such filters; The properties reported by *contact.engineering* are all computed on the *unfiltered* data that is only corrected using detrending and filling of missing values as described in section 3.2. For the detection of instrument noise and other artifacts, we apply a reliability cut-off based on tip-radius or instrument resolution. As for the separation of roughness and waviness, there is not a clear justification in the scientific literature for a sharp cutoff of size scales —therefore it is preferable to consider all of the multi-scale topography data in determining which scales matter, rather than making *a priori* assumptions. Second, there are distinctions in the way that slopes are calculated. For instance, SEMI MF1811 is inconsistent in proposing the forward differences (also shown in table 1) for a real-space calculation of $R_{\Delta q}$ but the Fourier derivative for the reciprocal space calculation of this quantity. ISO 4287 on the other hand proposes a sixth-order finite-difference scheme that will introduce significant smoothing at small scales. By contrast, the present web app computes these quantities in a way that avoids these inconsistencies, as shown in table 1.

Overall, while scalar parameters provide simple, clear descriptors, they must be interpreted with care. The RMS parameters of the *measurement* are not necessarily representative of the RMS parameters of the *physical surface* as represented by the digital surface twin. As shown in [37, 44, 45], the RMS parameters for a given measurement will vary by orders of magnitude depending on the scale at which they are measured. It can be useful to make relative comparisons of RMS parameters between topographies, as long as the measurements were performed in the same manner. However, the technique and size-range should always be explicitly specified along with the value. If it is instead desired to describe the parameters of the full *physical surface* rather than a *measurement*, then a stitched-together PSD may be used, with Parseval's law, to compute RMS parameters (as described in [20]). However, even these descriptors cannot reflect the size-dependent variation. In order to describe the height, slope, and curvature of a surface *as a function of size*, the scale-dependent roughness parameter (SDRP) analysis can be used, as described in the next section.

### 4.5. Scale-dependent roughness parameters (SDRP)

The purpose of the scale-dependent roughness parameters analysis is to explicitly show how the height, slope, and curvature vary as a function of the size scale that is being considered. We have described this approach in detail in [31], so only a brief description is included here.

The RMS parameters described in the previous section are all computed at the smallest relevant scale. For example, to compute the local slope, the discrete expression

$$(D_x h)_{m,n} \equiv \frac{h_{m+1,n} - h_{m,n}}{\Delta x} \tag{8}$$

considers the height difference between two adjacent pixels. We can generalize this derivative to

$$(D_x^{(\eta)} h)_{m,n} \equiv \frac{h_{m+\eta,n} - h_{m,n}}{\eta \Delta x} \tag{9}$$

which now considers the height difference between two pixels at distance $\ell = \eta \Delta x$. Note that the scale factor $\eta$ in equation (9) is an integer, but a similar expression can be defined for fractional scale factors, which however requires interpolation of the height values. We use linear interpolation for evaluating equation (9) at arbitrary $\ell$.

Using the scaled derivative to compute the RMS values summarized in table 2 yields the scale-dependent roughness parameters (SDRPs). Of particular interest is the scale-dependent RMS slope $h'_{\text{rms}}(\ell)$ and the scale-dependent RMS curvature $h''_{\text{rms}}(\ell)$. The true value of the underlying physical surface is the short distance limit $\ell \to 0$ of these properties, but care must be taken in interpreting this limit as it is most strongly affected by instrumental noise.

### 4.6. Power spectral density (PSD)

As discussed in detail in [20], there are many different mathematically valid forms of the PSD. The web application makes use of the conventions recommended in [20], and computes two different types of PSDs: the radially-averaged 2D-PSD $C^{\text{iso}}$; and the one-dimensional PSD $C^{1D}$ (referred to in [20] as $C^{1D+}$; however, the '+' is omitted here for simplicity). The 1D-PSD of a single line scan is given by $C_o^{1D} = L_x^{-1} |\tilde{h}_o|^2$, and the 2D-PSD of an area scan is given by $C_{o,p}^{2D} = (L_x L_y)^{-1} |\tilde{h}_{o,p}|^2$. We compute $C^{\text{iso}}(q)$ by (radially) averaging $C_{o,p}^{2D}$ over bins logarithmically spaced in the magnitude of the wavevector $q = |\vec{q}_{o,p}|$.

The 'correct' PSD to use for analysis of 2D measurements will depend on the surface, the measurement technique and parameters, and the desired application:

- The radially-averaged 2D-PSD $C^{\text{iso}}$ (units of m$^4$) is commonly used because it contains a full statistical description of the surface for an isotropic measurement. However, for a measurement that contains anisotropy due to instrumentation artifacts (as many AFM scans do), $C^{\text{iso}}$ will deviate from a 'true' description of the surface.

- The 1D-PSD $C^{1D}$ (units of m$^3$) is computed as the average of the individual PSDs computed from each line in a single direction. This is computed along both the *x*- and *y*-directions and the deviation between them serves as another measure of anisotropy (real or artifacted). For measurements that introduce artificial anisotropy (such as AFM), the $C^{1D}$ that is computed along the fast-scan direction may represent the most accurate possible description of the surface.

For comparisons between multiple surfaces, any of the above PSDs can be used and researchers should choose the most appropriate one based on anisotropy and ease of comparison. However, for quantitative calculations, such as the implementation of theoretical roughness models, a particular PSD must be used which is determined by the derivation of the model. For example, in the models developed by Persson (for example, [1, 46]), a modified version of $C^{\text{iso}}$ is required [20], $C^{\text{Persson}} = C^{\text{iso}}/4\pi^2$, to account for the differences in normalization between those models and the present calculations.

Radially-averaged 2D-PSDs can be converted into an equivalent one-dimensional (profile) representation via

$$\overline{C}^{1D}(q) \approx \frac{q}{\pi} C^{iso}(q) \qquad (10)$$

Note that this conversion is valid only where the PSD is self-affine; more information is given in [20]. To facilitate comparison between area scans and line scans, the web application displays both the pure 1D-PSDs $C^{1D}$ and the 2D-PSD $C^{iso}$ in the equivalent 1D representation (in units of m$^3$) as obtained from equation (10).

### 4.7. Height-difference autocorrelation function (ACF)

The height-difference autocorrelation function (ACF), also sometimes called the structure function [47], is the real-space equivalent of the PSD. It is formally defined as

$$A(\ell) = \frac{1}{2L} \int (h(x) - h(x + \ell))^2 dx \qquad (11)$$

i.e., as the correlation of the difference of heights between two points at distance $\ell$ on the surface. The ACF and PSD are connected by the Wiener-Khinchin theorem; one is the Fourier transform of the other. More information can be found in [25].

The ACF can be used to directly estimate some roughness parameters. For self-affine surfaces with Hurst exponent $H$, the ACF is a power-law $A(\ell) \propto \ell^{2H}$. The height-difference ACF furthermore saturates at the RMS height for distances where heights become uncorrelated, which we can formally write as $h_{rms}^2 = \lim_{\ell \to \infty} A(\ell)$. Dividing the square-root of the ACF by the distance yields the scale-dependent RMS slope of the surface, $h'_{rms}(\ell) = \sqrt{2A(\ell)}/\ell$. This is because equation (11) contains the first-order forward differences expression for the derivative (for more information see [31]). It also means that for small distances, the slope of the ACF corresponds to the RMS slope, $h'_{rms} \equiv h'_{rms}(0) = \lim_{\ell \to 0} \sqrt{2A(\ell)}/\ell$.

It is important to note that the web application displays the ACF on a double-logarithmic scale. The slope in this double-logarithmic scale is then proportional to the Hurst exponent $H$, not the RMS slope. The ACF needs to be plotted on a linear scale to estimate the slope. Alternatively, the slope can be estimated from the limiting value at small $\ell$ of the scale-dependent RMS slope described in the previous section. Like the PSD, the ACF can be reported for line scans, i.e., along certain directions, or as the radial average over all directions:

$$A(\ell) = \frac{1}{2\pi} \int \frac{1}{L_x L_y} \iint \frac{1}{2} (h(\vec{r}) - h(\vec{r} + \Delta \vec{r}(\ell, \phi)))^2 d^2 r d\phi \qquad (12)$$

with $\Delta \vec{r}(\ell, \phi) = (\ell \cos\phi, \ell \sin\phi)$. Unlike the PSD, the units for $A(\ell)$ are independent of whether we compute it for line scans, along certain directions, or as the radial average.

### 4.8. Variable bandwidth method (VBM)

While both ACF and PSD are commonly used to analyze the statistical properties of rough surfaces, *contact.engineering* also implements a slightly less common method for statistical analysis, the variable bandwidth method (VBM) [26–30]. The VBM directly encodes that, for a self-affine topography, the roughness parameters are intrinsically dependent on the lateral size of the measurement. The VBM displays the RMS height $h_{rms}$ as a function of the lateral size of the measurement, which we call the *bandwidth*. In the simplest incarnation, this is the size of the physical measurement, as for example demonstrated in [37, 44, 45].

The VBM approach can also be carried out on a single measurement. The measurement is subdivided into a checkerboard pattern (see inset to figure 2) of squares of equal lateral length $\ell = L/\zeta$ where we call $\zeta$ the *magnification*. We then compute $h_{rms}$ within each of these squares (after tilt correction in the individual square) and report the value $\langle h_{rms} \rangle$ that is obtained as the average over all squares. This procedure is repeated for magnifications that are an integer power of 2, i.e. $\zeta = 1, 2, 4, 8, \ldots$. An example of this procedure is shown in figure 2. The VBM can be thought of as a specific example of the more general SDRP analysis described above, though the implementation is slightly different (see [31] for more details.)

At the largest bandwidth ($\zeta = 1$), this simply yields the RMS height of the full measurement. For self-affine topographies, $\langle h_{rms}(\ell) \rangle \propto \ell^H$. The VBM has a simple intuitive interpretation in that it reports the width of the height distribution at different scales. While the one-dimensional PSD and ACF depend on the direction (or is reported as a radial average over different directions), the VBM has no directional dependence.

### 4.9. Reliability analysis

The SDRP, PSD, ACF, and VBM constitute four different methods that can be used to look at statistical properties of a measurement as a function of *scale* (denoted above by length $\ell$ or wavevector $q$). The measurement is accurate only over a portion of the full scale of the measurement. In particular, the smallest scales can be artifacted, because of instrumental noise, tip-radius artifacts, the diffraction limit, or other physical mechanisms that limit resolution. *Contact.engineering* has the capability to remove the inaccurate, artifacted portions of SDRP, PSD, ACF and VBM. This is of particular importance when stitching together individual measurements (see next section).

For optical profilometry, the user is asked to provide the resolution limit of the instrument. All data on length scales below this resolution limit is eliminated from the respective scale-dependent analysis.

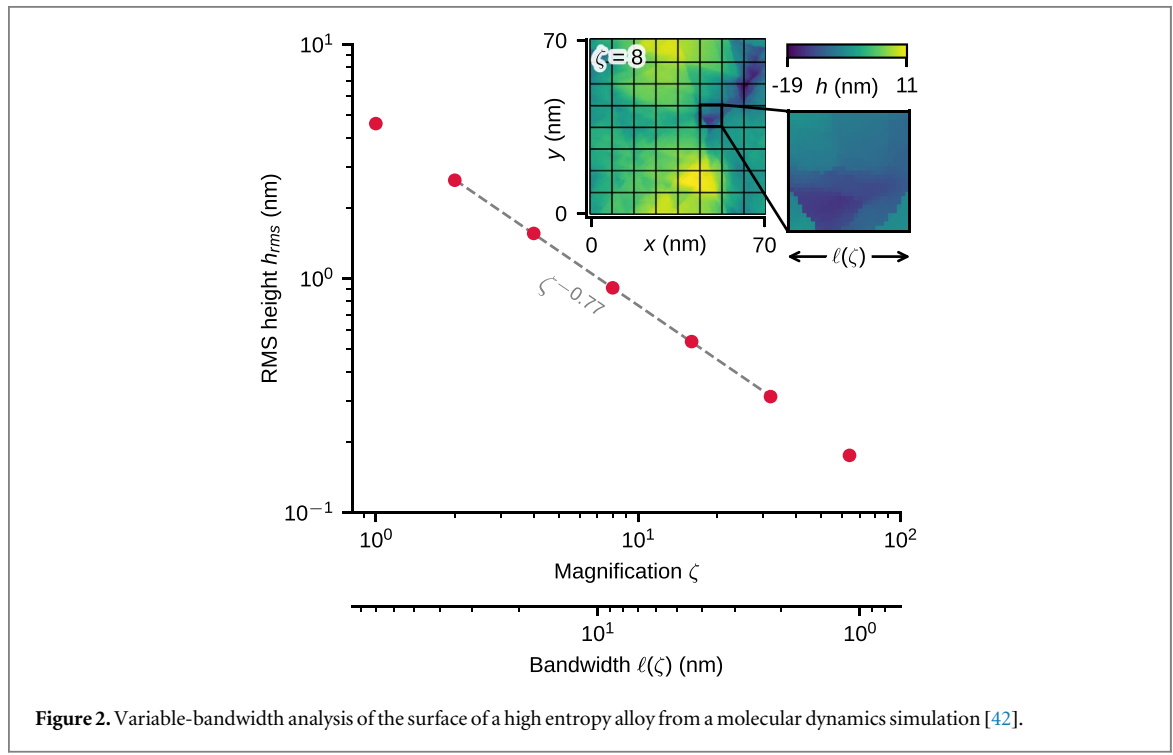

**Figure 2.** Variable-bandwidth analysis of the surface of a high entropy alloy from a molecular dynamics simulation [42].

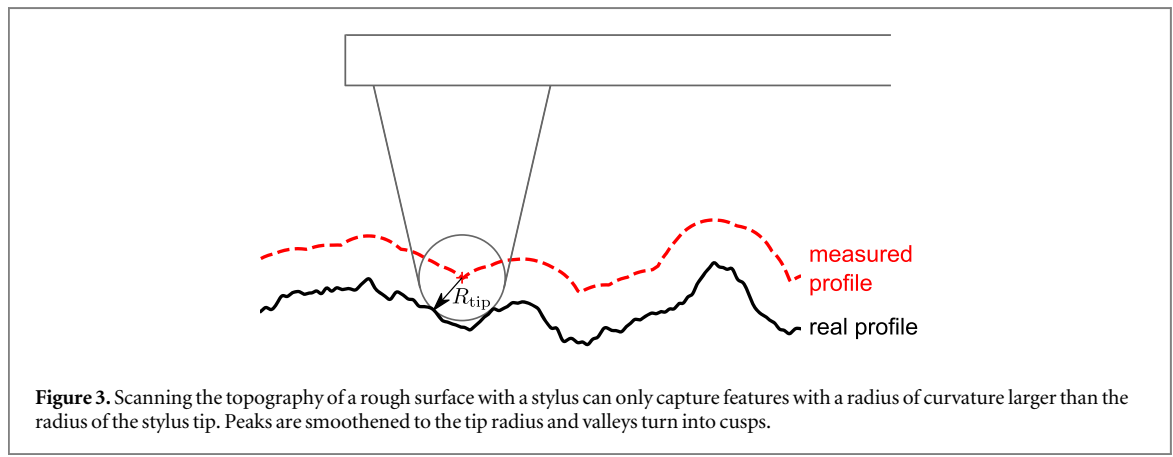

**Figure 3.** Scanning the topography of a rough surface with a stylus can only capture features with a radius of curvature larger than the radius of the stylus tip. Peaks are smoothened to the tip radius and valleys turn into cusps.

For tip-based profilometry, the user is asked to provide the tip radius $R_{\text{tip}}$ which is used to automatically detect tip artifacts. We use a tip radius analysis technique that originated in the work of Church & Takacs [48, 49] and was recently extended by the present authors [31]. Briefly, if the probe-tip radius $R_{\text{tip}}$ is larger than the radius of curvature of a valley, the scanning probe cannot fit into this valley on the rough topography. In this case, the peaks on the measured height data are rounded with a radius of curvature that cannot be smaller than the tip curvature. The valleys of the topography then develop cusps (figure 3). Church & Takacs pointed out [48] that the cusps lead to a (one-dimensional) PSD that scales as $C^{\text{1D}}(q) \propto q^{-4}$.

We have recently extended this idea [31], but we look at the rounded peaks instead of the cusps. We compute the local curvature at a certain scale $\ell$ via the scale-dependent second derivative, i.e., equation (9) or rather its equivalent for the second derivative. We then find the length $\ell$ below which the maximum of the (negative) curvature rises above the user-specified tip curvature,

$$\max_{x_m}\left[-\frac{D_{(\ell)}^2}{D_{(\ell)}x^2}h(x_m)\right] > \frac{c}{R_{\text{tip}}}. \qquad (13)$$

Any $\ell$ for which this condition is fulfilled must be unreliable, as a scanning probe with tip radius $R_{\text{tip}}$ is unable to scan regions with this curvature. The factor $c$ must be of order unity; we use $c = 1/2$ based on numerical experiments (see [31]). Criterion equation (13) hence adjusts the reliable region of the scale-dependent surface roughness data, with rougher surfaces generally leading to a larger unreliable region. Note that equation (13) expects peaks on the rough topography to have positive height values and care must be taken not to accidentally upload height data that has been flipped upside down.

### 4.10. The stitching together of surface descriptors: SDRP, PSD, ACF, and VBM

Almost all geometric properties of rough surfaces depend on scale, yet the existing engineering practice is to report only single values, such as $R_q$ or $S_q$. *contact.engineering* therefore promotes the use of SDRP, PSD, ACF, or VBM to report properties as a function of *scale*. The utility of reporting these properties as a function of scale is that measurements taken at different resolution, and even with different instruments, can be easily stitched together. The web application therefore displays the individual measurements of a surface in a single plot, exposing the underlying statistical structure of the physical surface.

The collection of all measurements of the same physical specimen constitute our digital surface twin. To obtain a single representation for SDRP, PSD, ACF, and VBM, we compute the averaged curves. Averages are taken over individual measurements after reliability analysis, i.e., portions of the measurement detected to be unreliable are not included in the average. The average is computed as the arithmetic mean over the datasets at specific logarithmically spaced collocation points. Note that we currently assign identical weights to each dataset, even though an individual line scan contains fewer points than an area scan. The underlying rationale is that consecutive lines on an area scan are correlated and it is presently unclear how to account for this when selecting weights for the averaging procedure.

For SDRPs, we compute the individual curves for all measurements at exactly the same collocation points. This requires interpolation of the line or area scan for the computation of the discrete derivative (see equation (9)) at fractional $\eta$. For PSD, ACF, and VBM we collect the measurements into logarithmically spaced bins over which we carry out the average. All averages are carried out over 1D (profile) representations of the respective analysis technique, since 2D representations are not available for line scans that may be part of the digital surface twin. In section 6 we will show examples of how this feature can be used to obtain a comprehensive description of the topography of a surface that goes beyond the individual measurement.

## 5. Contact calculations: Relationship between topography and surface properties, such as contact area, load, and normal displacement

### 5.1. Overview of common models to describe surface performance

A wide variety of models have been proposed to link the (elastic or plastic) mechanical deformation of rough topography to contact area or contact stiffness as intermediate properties from which other functional properties can be derived (e.g. [50]). These have been extensively reviewed elsewhere [1, 51–53]. In the broadest sense, they can be categorized into bearing-area models, independent-asperity models, renormalization group theory, and 'brute-force' numerical models. Bearing-area models assume that the surfaces in contact are rigid and treat contact as interpenetration of the rough topography; the contact area is then simply a slit-island analysis [54] of the rough topography. The resulting behavior of contact area versus penetration is often called the Abott-Firestone curve [55]. Independent-asperity contact models (such as the famous Greenwood and Williamson model [3] and its extensions, such as [56]) approximate roughness as a series of non-interacting hemispherical asperities. Persson's scaling theory [12, 13] and the approach by Joe, Thouless, & Barber [57] solve the contact problem at a specified scale given that the solution at larger scales is known, leading to a renormalization of contact properties. Bearing-area, independent-asperity and renormalization models require as input statistical descriptions of the topography in the form of RMS parameters, the power spectral density, or others. By contrast, 'brute-force' numerical calculation solve for the elastic deformation using a *specific* realization of topography [16, 53, 58–78]. While these calculations require fewer simplifications than the other categories and so may result in more accurate predictions, they are resource-intensive to run and do not typically yield simple equations that are generalizable to other surfaces with similar statistics.

All four categories of rough-surface models have been proven successful in certain contexts, but all are based on simplifying assumptions, and it remains difficult to experimentally validate their predictions and to directly apply them to real-world surfaces to optimize roughness for a desired surface property. At present, the web application focuses primarily on the fourth class of models: the numerical implementation of continuum contact mechanics. However, additional computational workflows are constantly being added.

### 5.2. Assumptions behind the boundary element method (BEM)

*Contact.engineering* computes exactly the elastic deformation of two frictionless contacting topographies for a linear-elastic, isotropic solid, and is able to consider plasticity using approximate models. The geometry considered is that of a rigid rough surface on an elastic half-space, and we only consider the displacements of the surface in the normal direction. For frictionless contact, the only error introduced by the latter approximation is that asperities can be displaced within the plane of contact, leading to slightly different contact geometries. This latter approximation is commonly employed in state-of-the-art contact calculations [16, 53, 58–78] and allows to map the contact of any two elastic solids with arbitrary geometry onto that of a rigid rough surface on an elastic flat surface. This approach is exact for the contact of two elastic solids with Poisson ratio $\nu = 1/2$. We note that consideration of the relative motion of the two solids during contact, introduced by differences in the elastic moduli, would require frictional constitutive laws, significantly complicating the analysis.

All results with units of pressure (such as normal pressure or normal force) are expressed in terms of the effective contact modulus $E^*$, with $1/E^* = (1 - \nu_1^2)/E_1 + (1 - \nu_2^2)/E_2$, where $E_1$ and $E_2$ are the Young's moduli of the two contacting bodies and $\nu_1$ and $\nu_2$ their Poisson ratios [79]. Conversion to real pressure units is therefore the responsibility of the user. For plastic calculations, as discussed below, there is then only a single material parameter $p_Y/E^*$, the ratio of hardness $p_Y$ to the modulus $E^*$. For the contact of two rough surfaces, the user is at present required to take a measurement of each surface, $h_1(x, y)$ and $h_2(x, y)$, and then to create a compound topography $h_1(x, y) + h_2(x, y)$ for upload. We note that the analysis is carried out for individual measurements or compound topographies and cannot be averaged to be representative for a full digital surface twin. Contact properties also depend on the scale of the measurement and care has to be taken in their interpretation. In particular, contact area depends on small-scale properties (in particular $h'_{\rm rms}$) [11, 14, 80] while contact stiffness depends on the largest scale (in particular $h_{\rm rms}$) [68, 81, 82].

### 5.3. Solution of the elastic problem

The elastic deformation of the substrate is computed using a boundary element method that considers just the degrees of freedom of the surface and treats the bulk as an elastic half-space. Such boundary element methods have been extensively described in the literature. Examples include direct summation of the (regularized) Boussinesq-Cerutti solution [79] for point load or multilevel summation techniques [60]. We solve the elastic problem in reciprocal space using a fast Fourier transform (FFT) technique, that is similar to the following: the approach of Stanley & Kato [59]; the DC-FFT technique [83]; and related techniques for atomic lattices [84–86].

The FFT can be used to accelerate the computation of a convolution in real-space. In linear elasticity, the surface displacements $u(x, y)$ are related to the surface pressure $p(x, y)$ through a linear operator, the elastic surface Green's function $G(x, y)$:

$$u(x, y) = \int_{A_G} G(x - x', y - y') p(x', y') dx' dy'$$
$$= G(x, y) * p(x, y) \quad (14)$$

The Fourier transform of equation (14) turns the convolution $G * p$ into a simple product

$$\tilde{u}(q, k) = \tilde{G}(q, k) \tilde{p}(q, k), \quad (15)$$

where a tilde $\tilde{f}$ indicates the Fourier transform of a function $f$:

$$\tilde{f}(q, k) = \int_{A_G} f(x, y) e^{-iqx - iky} dx dy \quad (16)$$

Note that the integral is carried out over the area of support of the Green's function, $A_G$. For nonperiodic calculations, this differs from the nominal area of the measurement $A_0$ by a factor of 4, since the computation requires a padding region as described below.

The Green's function for the elastic problem can be analytically or semi-analytically derived for periodic [84–88] and non-periodic [58, 89, 90] systems in the continuum limit [87, 89, 90], for atomic lattices [84–86] and for substrates of finite thickness [91–94]. The web service uses the continuum expressions and can handle periodic and non-periodic systems.

The linear-elastic Green's function for periodic pressure distributions takes the particularly simple form in reciprocal (Fourier) space [88]

$$\tilde{G}(\vec{q}) = 2/(E^*|\vec{q}|). \quad (17)$$

Surface topographies are typically provided as measurements of rectangular sections of size $L_x \times L_y$, measured on a homogeneous grid of $N_x \times N_y$ pixels. The nominal measurement resolution is then $\Delta x = L_x/N_x$ and $\Delta y = L_y/N_y$. (The true resolution can of course be lower because of instrumental artifacts, see discussion above.) Data and calculations are therefore provided on a discrete grid. The discrete version of equation (14) replaces the integral with a sum. Equation (16) then becomes the discrete Fourier transform, which we compute with a standard FFT algorithm on the same grid as the uploaded surface. This discretization limits the range of wavevectors that enter the periodic Green's function, equation (17), to $\tilde{G}_{mn} = \tilde{G}(2\pi m/L_x, 2\pi n/L_y)$ with $m \in [0, 1, \ldots, \lfloor(N_x - 1)/2\rfloor, -\lfloor N_x/2 \rfloor, \ldots, -1]$ and $n \in [0, 1, \ldots, \lfloor(N_y - 1)/2\rfloor, -\lfloor N_y/2 \rfloor, \ldots, -1]$.

The situation for nonperiodic contacts is more complicated. First, we can no longer straightforwardly use equation (17) for the Green's function because this would require a numerical grid of infinite size to decouple images. Carrying out the inverse Fourier transform of equation (17) analytically for this case yields the Boussinesq-Cerutti functions that describe the response of the elastic solid to a concentrated normal force [79],

$$G_{\rm BC}(\vec{r}) = 1/(\pi E^* |\vec{r}|). \quad (18)$$

This function diverges as $r \to 0$ and must be regularized. The typical approach is to distribute the contacting force over the area associated with the discrete surface element (of size $\Delta x \times \Delta y$) and assume constant pressure on that element [58]. Our elements are rectangles and we use the expression for constant pressure on a rectangular surface area [79, 95]. We note that more sophisticated schemes such as linear or quadratic interpolation of pressure between discretization points have been described in the literature [58, 89, 96], but practical differences only occur at small scales where other effects (e.g. measurement artifacts) dominate.

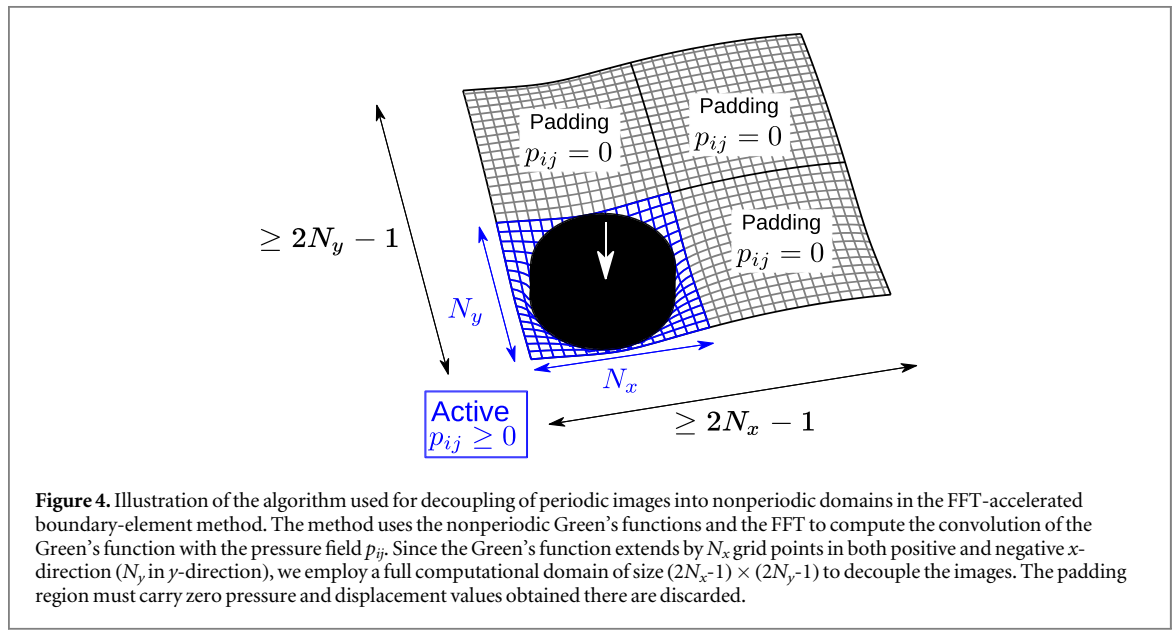

**Figure 4.** Illustration of the algorithm used for decoupling of periodic images into nonperiodic domains in the FFT-accelerated boundary-element method. The method uses the nonperiodic Green's functions and the FFT to compute the convolution of the Green's function with the pressure field $p_{ij}$. Since the Green's function extends by $N_x$ grid points in both positive and negative $x$-direction ($N_y$ in $y$-direction), we employ a full computational domain of size $(2N_x-1) \times (2N_y-1)$ to decouple the images. The padding region must carry zero pressure and displacement values obtained there are discarded.

The final Green's function for nonperiodic systems that *contact.engineering* uses is [79]

$$\pi E^* G(x, y)$$
$$= (x + a) \ln \left[ \frac{(y + b) + \{(x + a)^2 + (y + b)^2\}^{1/2}}{(y - b) + \{(x + a)^2 + (y - b)^2\}^{1/2}} \right]$$
$$+ (y + b) \ln \left[ \frac{(x + a) + \{(x + a)^2 + (y + b)^2\}^{1/2}}{(x - a) + \{(x - a)^2 + (y + b)^2\}^{1/2}} \right]$$
$$+ (x - a) \ln \left[ \frac{(y - b) + \{(x - a)^2 + (y - b)^2\}^{1/2}}{(y + b) + \{(x - a)^2 + (y + b)^2\}^{1/2}} \right]$$
$$+ (y - b) \ln \left[ \frac{(x - a) + \{(x - a)^2 + (y - b)^2\}^{1/2}}{(x + a) + \{(x + a)^2 + (y - b)^2\}^{1/2}} \right]$$
(19)

where $a = \Delta x/2$ and $b = \Delta y/2$ are half the grid spacing in $x$- and $y$-directions.

The use of the nonperiodic real-space Green's function, equation (19), with the FFT convolution technique, equation (15), requires an additional trick to decouple periodic images. This trick was described by Hockney in the context of the solution of the Poisson equation for electrostatic problems [97] and later employed, e.g., to decouple images in plane-wave density functional calculations [98]. In contact mechanics, this method was introduced by multiple authors to decouple periodic images [73, 83]. In brief (see also figure 4), we solve the equation (14) on a grid of at least $(2N_x - 1) \times (2N_y - 1)$ grid points (and area $A_G$) and require that all pressures $p_{ij} \equiv 0$ for $i \geqslant N_x$ or $j \geqslant N_y$. The reciprocal space Green's function is then obtained by discretizing equation (19) on a regular grid, $G_{mn} = G(\Delta x m, \Delta y n)$ with $m \in [0,1,\ldots,N_x - 1, -(N_x - 1), \ldots, -1]$ and $n \in [0,1,\ldots,N_y - 1, -(N_y-1), \ldots, -1]$, and then computing the discrete Fourier transform of $G_{mn}$. Note that this is the minimum grid size required to decouple the periodic images. Large grid sizes are possible and sometimes beneficial for optimal FFT performance.

The regions with $p_{ij} \equiv 0$ are padding regions that decouple the images (see figure 4). This works because the Green's function $G_{kl}$ has a maximum range of $N_x - 1$ in $x$- and $N_y - 1$ in $y$-direction. Because of the padding region, any two points in the 'active' region (see figure 4) are at most $(N_x - 1, N_y - 1)$ apart and any distance vector crossing the padding region is longer than $(N_x - 1, N_y - 1)$. Note that displacements $u_{kl}$ within the padding region have no physical meaning: They come from a superposition of repeating images and must be ignored. The web app returns just pressure and displacements in the active region with area $A_0$; the inactive padding region is hidden from the user.

### 5.4. Solution of the contact problem

The previous section discussed the calculation of the relationship between local pressure and elastic displacement of the substrate's surface. We use an FFT-based algorithm, but this FFT algorithm can in principle be replaced by any other (equivalent) solution of this problem described in the literature, such as direct summation [58] or multilevel summation [60]. In our experience, the FFT-based formulation yields the best numerical performance across application scenarios.

The second part to any solution of a contact problem is the contact condition itself. The continuum mechanics picture, which is also the one employed at present in *contact.engineering*, is that of two impenetrable solids. One of those is flat and deformable. When not contacted, it is located at $u_{ij} \equiv 0$. The counterbody is described by the discrete function $h_{ij}$. In mathematical terms, the contact condition then becomes a linear complementarity problem [58]:

$$p_{ij} \geqslant 0, \ u_{ij} - h_{ij} \geqslant 0, \ p_{ij}(u_{ij} - h_{ij}) = 0. \quad (20)$$

We use the constrained conjugate gradient algorithm of Polonsky & Keer [60] to solve equation (20). The variable in Polonsky & Keer's algorithm is the pressure $p_{ij}$. The conjugate gradient scheme optimizes $p_{ij}$ so that the overlap $h_{ij} - u_{ij}$ is close to zero within the contacting regions $I_c = \{(i, j) | p_{ij} > 0\}$. After each step, the algorithm constrains $p_{ij} \geqslant 0$ and whenever a point outside the contact region (defined by $p_{ij} = 0$) enters into contact ($u_{ij} \leqslant h_{ij}$), the conjugate gradient iteration is reset. More information on this algorithm can be found in [60]. Our implementation of this algorithm also allows to specify an upper limit to the pressure on the surface, corresponding to an indentation hardness $p_y$ [62, 99].

We note that the algorithm is fast at low contact area but requires many iterations to converge at large contact area. By default, we stop the algorithm after 100 iterations. These non-converging calculations are reported to the user but the dataset is shown with translucent markers to make the user aware that the data may be unreliable. The user can change the number of iterations up to a hard upper limit of 1000.

## 6. Examples of use

In the following section, we present two examples of how to use *contact.engineering*. We note that all figures presented here are reproduced as shown by the web service to illustrate that publication-ready figures can be directly obtained from *contact.engineering*. The first example illustrates the use of statistical analysis techniques to stitch together many measurements, while the second example illustrates contact calculations on individual topographies.

### 6.1. Ultrananocrystalline diamond
In order to illustrate that many measurements can be stitched together to arrive at a statistical representation of a rough surface, we analyze the data of more than 100 measurements of an ultrananocrystalline diamond film reported in [37, 45]. The dataset contains topography data obtained from stylus profilometry, atomic force microscopy (AFM), and also using a less-common microscopy approach: sideviews and cross-sections in a transmission electron microscope (TEM). The TEM and stylus profilometry approaches yield line scans while the AFM yields area scans. This illustrates that it is possible to combine and analyze structurally distinct data within a single digital surface twin. We note that we are repeating the analysis of [31, 37, 45]; this shows that anyone can repeat this sophisticated analysis without the need for complex mathematical calculations or specialized knowledge.

The whole dataset was uploaded to create a digital surface twin, and then analyzed using the PSD, ACF, VBM, and scale-dependent curvature, as shown in figure 5. All data has been tilt-corrected before analysis. Data from the scanning probe techniques (AFM and stylus) has been filtered to remove unreliable portions (see discussion on tip artifacts in section 4.9); the remaining dataset contains only data deemed reliable and the averages (thick black curves) are only computed over this reliable dataset. The plot illustrates that the individual measurements of all four methods line up and can be used to obtain a statistical representation of the surface.

All analysis methods show a region of power-law scaling, which occurs at large wavevectors for the PSD (figure 5(a)) and at short distances or bandwidths for ACF, VBM, and SDRP (figures 5(b)–(d)). The VBM (figure 5(c)) saturates at the RMS height of the surface of roughly 10 nm. The ACF saturates at the square of this value. The SDRP shows that the curvature is a function of scale that appears to grow unbounded as resolution increases (distance decreases). Note that the curvature shown in figure 5(d) cannot be straightforwardly extracted from the other plots, showing that the SDRP is useful to obtain parameters that are straightforward to interpret geometrically but difficult to obtain by other means. The ACF (figure 5(b)) shows larger deviation from the average, e.g., at distances in the range 10-100 nm. Those are artifacts introduced by tilt-correcting individual measurements: Tilt correction forces the ACF to go to zero at the size of the measurement, resulting in a downtick of the ACF curve. None of the other techniques is sensitive to this artifact.

To illustrate the advancement of FAIR data practices, we note that this digital surface twin of the ultrananocrystalline diamond has been published and is DOI-accessible at [100]. This means that anyone can access the surface, including all of the underlying raw topography data, for examination or re-use. Furthermore, all of the above analysis can be repeated, and any other desired analysis can be run, by anyone who accesses the site.

### 6.2. Smooth- and rough-sphere contacts
As an example of the use of elastic contact-mechanics simulations, we simulate the contact of a rough sphere. This digital surface twin is not based on real-world measurements but on computer-generated topography, defined on a 1024-by-1024 grid with a pixel size of 1 nm by 1 nm. Elasticity was computed using free boundaries as described in section 5. The sphere was generated by superposing a nominally flat self-affine random field generated by Fourier-synthesis [20, 39] and a paraboloid.

The relationship between contact area and load (mean pressure) is shown in figure 6(a), as displayed by *contact.engineering*. For reference, we also show contact areas for the smooth paraboloid and for the roughness alone, where we used periodic boundary conditions for the nominally flat roughness. It can be observed that the contact-area-versus-pressure curve corresponds to the rough surface at low loads but crosses over to Hertzian for high pressures, as described in [73]. For the periodic nominally flat roughness at high contact ratios, some simulations did not reach convergence in the number of allowed iterations (1000). These simulations are indicated by a translucent datapoint.

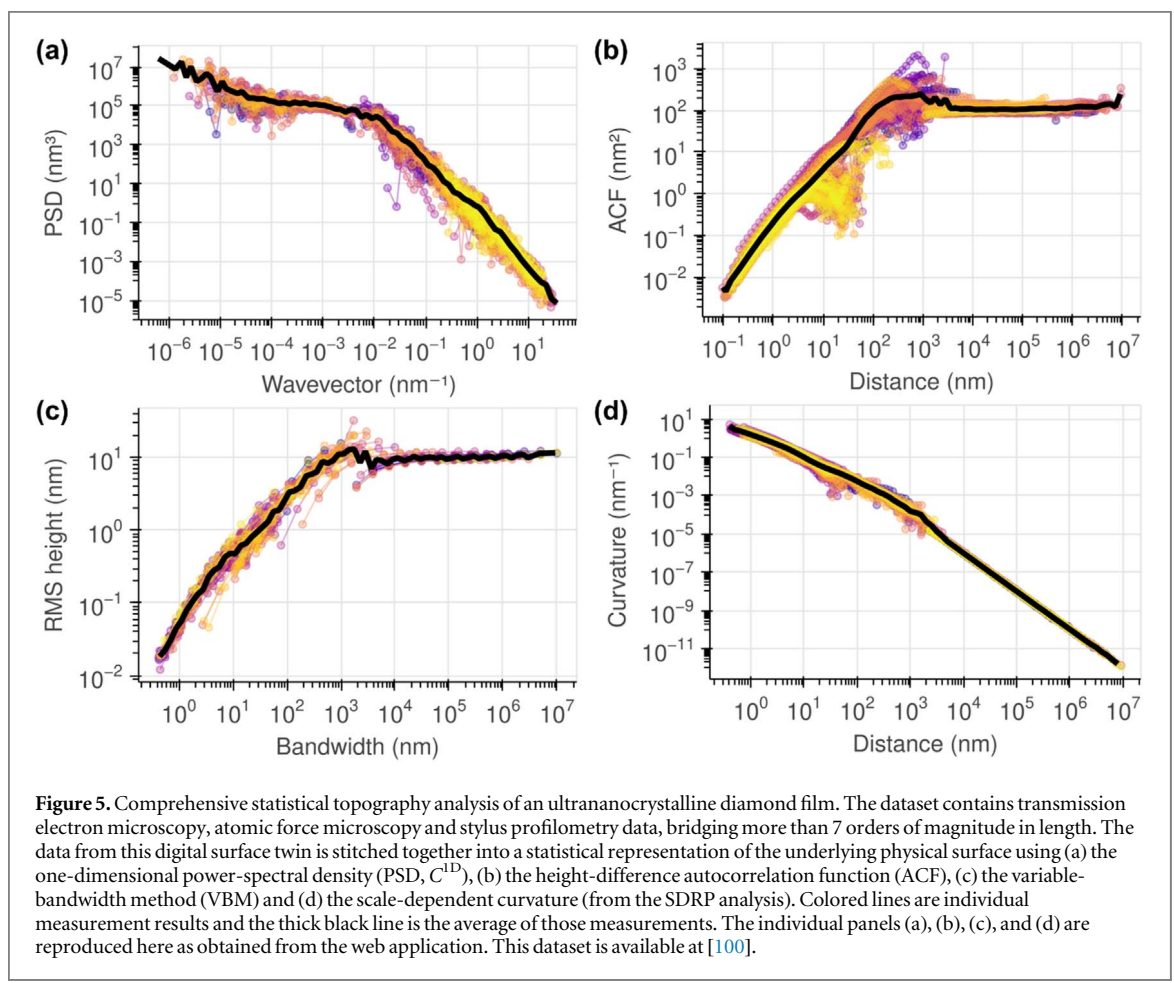

**Figure 5.** Comprehensive statistical topography analysis of an ultrananocrystalline diamond film. The dataset contains transmission electron microscopy, atomic force microscopy and stylus profilometry data, bridging more than 7 orders of magnitude in length. The data from this digital surface twin is stitched together into a statistical representation of the underlying physical surface using (a) the one-dimensional power-spectral density (PSD, $C^{1D}$), (b) the height-difference autocorrelation function (ACF), (c) the variable-bandwidth method (VBM) and (d) the scale-dependent curvature (from the SDRP analysis). Colored lines are individual measurement results and the thick black line is the average of those measurements. The individual panels (a), (b), (c), and (d) are reproduced here as obtained from the web application. This dataset is available at [100].

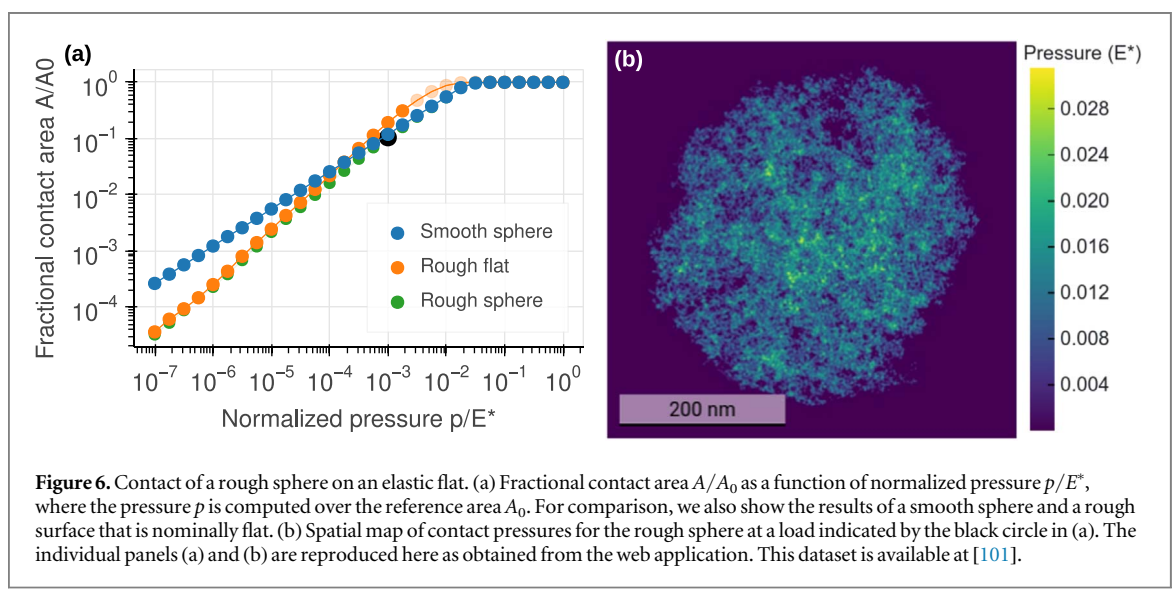

**Figure 6.** Contact of a rough sphere on an elastic flat. (a) Fractional contact area $A/A_0$ as a function of normalized pressure $p/E^*$, where the pressure $p$ is computed over the reference area $A_0$. For comparison, we also show the results of a smooth sphere and a rough surface that is nominally flat. (b) Spatial map of contact pressures for the rough sphere at a load indicated by the black circle in (a). The individual panels (a) and (b) are reproduced here as obtained from the web application. This dataset is available at [101].

In addition to the plot of contact area versus load, *contact.engineering* allows the visualization of a map of the contact pressure for each simulation (figure 6(b)), as well as the gap distribution and other quantities. The contact pressure reflects the rough topography of the sphere, but appears highest in the center and lowest as the edges as would be predicted by Hertz theory.

The specific pressure map of figure 6(b) is obtained right where the rough sphere crosses over from flat-on-flat to Hertz behavior and is indicated by the black circle in figure 6(a). The pressure is more homogeneous below this crossover (lower loads) and becomes more Hertzian-like at higher loads. More details can be found in [73].

## 7. Software implementation and infrastructure

We here briefly comment on the software implementation of the methods described here. All software is published under the MIT license and available at https://github.com/ContactEngineering and we encourage engagement from the scientific community. We follow established practices in development of scientific software [102], such as continuous integration with an appropriate automated testing framework to ensure software quality.

We distinguish between back-end and front-end modules. Back-end modules handle the topography data and carry out numerical analyses. The package *SurfaceTopography* implements filters for reading topography measurement data and performing statistical analysis (PSD, ACF, etc.) on it. It also has provisions for computing averages over multiple measurements. The package *ContactMechanics* implements boundary-element calculations for obtaining contact area, pressure, and other associated quantities. These back-end codes use *numpy* [103] for numerical calculations and implement all the scientific, algorithmic functionality described in this paper. The back-end is parallelized using the message passing interface (MPI) and can be used independently of the web front, for example on high-performance computing systems.

The web front end *TopoBank* is based on *Django* that provides bookkeeping of uploaded data via interaction with an underlying *PostgreSQL* database and a storage system, currently a *NetApp StorageGRID* instance running georedundantly in Freiburg and Tübingen, Germany. *Django* also renders HTML pages and handles user interaction. *TopoBank* is split into a manager that handles upload and visualization of measurements and digital surface data (including their metadata), and an analysis module that orchestrates running pre-defined analyses and visualizing them. Analyses themselves are short functions that call the respective analysis methods in *SurfaceTopography* or *ContactMechanics* and place results in the storage system. Analysis functions are distributed among compute nodes and prioritized via the *Celery* task queue. All analyses are computed asynchronously, i.e., they are placed in the *Celery* task queue and become available for visualization once they have run. This means that once they have run, the results are instantly available to the user for visualization. Some analyses allow specification of parameters. Changing a parameter only triggers a new calculation if no calculation with the same parameter set has already run. Calculations with a default parameter set run immediately after upload for all analyses.

## 8. Summary and conclusions

We have described a web-based application that is designed to standardize the analysis of topography data and the calculation of surface performance. The *contact.engineering* application addresses the three central challenges of surface analysis, which are described in the Introduction section. First, the application puts a focus on collecting many measurements for the surface of a real-world specimen or device, at various length scales and using different instruments, and integrating them into a digital surface twin of the real-world incarnation. The measurements are stitched together to describe all measured scales of the physical surface. The more measurements that are uploaded and the wider the variety of length scales, the more accurate the representation of the digital surface twin, with the goal of a *comprehensive* description of multi-scale surface topography. Second, the web application implements advanced analysis tools, including statistical metrics to describe the comprehensive surface topography, as well as mechanical models to predict the surface performance. Third, the application allows users to securely store, share, and optionally publish the digital surface twins and all associated analysis, in a manner that implements a FAIR data policy [35]. The published datasets receive digital object identifiers (DOIs) and can be cited in publications, in reports to funding agencies, or even in advertising a new material or technique. We hope that the service will become a central repository of surface topography data for scientific research and industry applications.

This document describes the current state of the service; we are constantly evolving it towards better usability and more analysis features. Right now, the *digital surface twin* only contains topography data, but in the future will incorporate an even broader set of surface characterizations, such as chemical composition, as well as computational models for the evolution of roughness [42, 43]. We will also work towards incorporating standardized ontologies that are currently being developed for tribological problems [36]. The overarching goal is to get ever closer to a complete digital surface twin that accurately reflects surface performance.

To aid in this process, we would like to encourage suggestions from the scientific and industry communities about ways to further improve data standardization, data analysis, and the advancement of understanding and application of surface finish as a means to improve surface performance. A starting point is the discussion forum on the GitHub repository at https://github.com/ContactEngineering/TopoBank/discussions or by email to the authors of this article. The web service itself is free of charge to the scientific community and accessible at https://contact.engineering/.

## Acknowledgments

This article is dedicated, and the authors are deeply indebted, to the late Mark O. Robbins, who was involved in developing the basic ideas behind *contact.engineering* and starting the project. We thank Tobias Amann, Roland Bennewitz, Claas Bierwisch, Sitangshu Chatterjee, Michele Ciavarella, Qinglin Deng, Martin Dienwiebel, Robert Jackson, Tobias Kling, Mena Klittich, Richard Leute, Zhuohan Li, Gianpietro Moras, Michael Moseler, Nathaniel Orndorf, Antonio Papangelo, Chris Thom and Yang Xu for patience in testing early versions of *contact.engineering*, reporting bugs and suggesting changes to interface and functionality. We are indebted to Johannes Hörmann, Jan Leendertse and Dirk von Suchodoletz for enlightening discussions on open data and research data management, Kolja Glogowski and Marcel Tschöpe for help with the storage infrastructure, and Saher Seeman and Jannis Seyfried for enabling DOI registration. We thank Mathys AG Bettlach (Switzerland) for providing the image of their hip implant and Michal Rössler for designing the contact.engineering logo, both shown in figure 1. The work presented in this paper is the joint outcome of multiple projects, funded by the European Research Council (StG 757343), the Deutsche Forschungsgemeinschaft (grants PA 2023/2, EXC 2193/1—390951807 and INST 39/1099-1 FUGG) and the U.S. National Science Foundation (CMMI-1727378 and CMMI-1844739).

## Data availability statement

The data that support the findings of this study are openly available at the following URL/DOI: https://doi.org/10.57703/ce-5cz7a, https://doi.org/10.57703/ce-8ppx5.

## ORCID iDs

Michael C Röttger  https://orcid.org/0000-0002-7111-8570
Antoine Sanner  https://orcid.org/0000-0002-7019-2103
Luke A Thimons  https://orcid.org/0000-0003-4511-1807
Till Junge  https://orcid.org/0000-0001-8188-9363
Wolfram G Nöhring  https://orcid.org/0000-0003-4203-755X
Tevis D B Jacobs  https://orcid.org/0000-0001-8576-914X
Lars Pastewka  https://orcid.org/0000-0001-8351-7336